\documentstyle[12pt]{article}
\newcommand{\slq}{\raise.15ex\hbox{$/$}\kern-.57em\hbox{$q$}}
\newcommand{\slp}{\raise.15ex\hbox{$/$}\kern-.57em\hbox{$p$}}
\newcommand{\be}{\begin{equation}}
\newcommand{\ee}{\end{equation}}
\newcommand{\ba}{\begin{eqnarray}}
\newcommand{\ea}{\end{eqnarray}}

\newcommand{\Tr}{{\rm{Tr}}}

\date{}

\newcommand{\la}{\langle}
\newcommand{\ra}{\rangle}
\newcommand{\MeV}{{\rm MeV}}
\begin{document}
\begin{titlepage}
\begin{flushright}
HD--THEP--96--24\\
Fermilab--Pub--96/194--T\\
July, 1996
\end{flushright}
\quad\\
\vspace{0.8cm}
\begin{center}
{\bf\LARGE Quark Masses}\\
\medskip
{\bf\LARGE from the Linear Meson Model}\\
\vspace{1cm}
Dirk--Uwe Jungnickel\footnote{Email:
D.Jungnickel@thphys.uni-heidelberg.de} and Christof
Wetterich\footnote{Email: C.Wetterich@thphys.uni-heidelberg.de}\\
\bigskip
Institut  f\"ur Theoretische Physik\\
Universit\"at Heidelberg\\
Philosophenweg 16, D-69120 Heidelberg\\
\vspace{1cm}
\end{center}
\begin{abstract}
Quark mass ratios are expressed within the linear meson
model by universal relations involving only the masses
and decay constants of the flavored pseudoscalars as well as
their wave function renormalization. Quantitative results
are in agreement with those obtained from chiral perturbation
theory, with a tendency to a somewhat higher strange
quark mass.
\end{abstract}\end{titlepage}
\newpage
Chiral perturbation theory can predict the ratios of (current) quark
masses with a very satisfactory accuracy \cite{1}-\cite{3}.
Nevertheless, some assumptions about the convergence of the quark mass
expansion have to be made since corrections quadratic in the quark
masses are usually neglected. Even though these assumptions are quite
reasonable it would be valuable to have an independent check of their
validity. This can be provided within a linear meson model. In this
model a complex $3\times 3$ matrix $\Phi$ describes simultaneously the
pseudoscalar $(O^{-+})$ octet and singlet as well as the scalar
$(O^{++})$ octet and singlet. The meson decay constants are related to
the expectation values of the unflavored scalars which constitute the
real diagonal part of $\Phi$. In presence of quark masses these
expectation values also determine the flavored meson masses \cite{4}.
One therefore expects relations between the masses and decay constants
of the flavored mesons and the quark masses. We essentially exploit
only symmetry properties and work within the framework of an effective
action. This generates the 1PI-Green functions and all quantum
fluctuations are supposed to be included in the effective coupling
constants.

We start with the most  general form of the effective action consistent
with the flavor symmetry $SU_L(3)\times SU_R(3)$ as well as
parity and charge conjugation
\be\label{1}
\Gamma[\Phi]=\int d^4 x\left\{{\cal L}_{\rm kin}+U-
\frac{1}{2}\Tr(\Phi^\dagger j+j^\dagger\Phi)\right\}.\ee
Here the kinetic term ${\cal L}_{\rm kin}$ contains all terms
involving derivatives of $\Phi$, $U$ is the effective potential and
the source term describes the response to non-vanishing quark
masses. We work here within a formalism where $\Phi$ represents a
composite field for quark--antiquark states and all flavor symmetry
breaking is cast in a linear coupling of $\Phi$ to the source term. We
consider real and diagonal sources
\ba\label{2}
j&=&{\rm diag} (j_u,j_d,j_s)\nonumber\\
j_q&=&2C  m_q\ea
where the current quark masses $m_q$ are evaluated at a convenient
scale (say in the $\overline{MS}$ scheme at $\mu=1$ GeV). The minimum of $U-
\frac{1}{2}\Tr(\Phi+\Phi^\dagger)j$ occurs for real and diagonal $\Phi$,
\be\label{3}
\la\Phi\ra={\rm diag} (\la\varphi_u\ra,\la\varphi_d\ra,
\la\varphi_s\ra)\ee
such that the discrete symmetries $C$ and $P$
remain conserved. For a given form of the effective potential the
expectation values are determined by the field equations
for the real diagonal components of $\Phi$
\be\label{4}
\frac{\partial U}{\partial\varphi_q}_{|\la\Phi\ra}=j_q.\ee
In turn, the  unrenormalized mass matrix can now be inferred from the
second derivatives of $U$ evaluated for $\Phi=\la\Phi\ra$.
 We will denote the eigenvalues for the flavored pseudoscalars by
$\overline M^2_{\pi^{\pm}},\overline M^2_{K^\pm}$ and
$\overline M^2_{K^0}$.

In order to connect the (zero momentum) mass terms $\overline M^2_i$
to the physical pole masses $M^2_i$ one also needs information
contained in the kinetic terms. For the flavored pseudoscalars
there is no mixing and we can always write the most general momentum
dependence of their inverse propagators as
\be\label{5}
G^{-1}_i(q)=\overline M^2_i+Z_i q^2+h_i(q^2).
\ee 
Here the normalization conditions for $\overline M^2_i$ and $Z_i$ are
formulated as $h_i(-M_i^2)=h_i(0)=0$ \cite{4} implying the simple
relation
\be\label{6}
M^2_i=\overline M^2_i/Z_i.
\ee
The decay constants of the flavored pseudoscalars are defined by
their leptonic decays and can again be expressed \cite{4} in terms
of $\la\varphi_q\ra$ and $Z_i$
\begin{eqnarray}
\label{7}
f_\pi&=& Z_\pi^{\frac{1}{2}}(\la\varphi_u\ra+\la\varphi_d\ra)
=Z_\pi^{\frac{1}{2}}\bar f_\pi\nonumber\\
f_{K^\pm}&=&Z^{\frac{1}{2}}_{K^\pm}(\la\varphi_u\ra+\la\varphi_s\ra)
=Z_{K^\pm}^{\frac{1}{2}}\bar f_{K^\pm}\nonumber\\
f_{K^0}&=&Z^{\frac{1}{2}}_{K^0}(\la\varphi_d\ra+\la\varphi_s\ra)
=Z_{K^0}^{\frac{1}{2}}\bar f_{K^0}.
\end{eqnarray}
For a given effective potential, say, for example,
\begin{eqnarray}
\label{8}
U&=&\overline m_g^2(\rho-3\bar\sigma_0^2)-\frac{1}{2}\bar \nu(\xi-\bar
\sigma_0 \rho+\bar\sigma_0^3)
+\frac{1}{2}\bar\lambda_1(\rho-3\bar\sigma_0^2)^2+\frac{1}{2}
\bar\lambda_2\tau_2+\frac{1}{2}\bar\lambda_3\tau_3\nonumber\\
\rho&=&\Tr \Phi^+\Phi,\quad \tau_2=\frac{3}{2}\Tr
(\Phi^+\Phi-\frac{1}{3}\rho)^2\nonumber\\
\tau_3&=&\Tr(\Phi^+\Phi-\frac{1}{3}\rho)^3,\qquad \xi=\det \Phi+\det
\Phi^+
\end{eqnarray}
one can now relate $\overline M^2_i$ and $\bar f_i$ to the quark masses.
Hereby the field equations (\ref{4}) and the expressions for
$\overline M^2_i$ become rather lengthy expressions involving the
parameters $\overline m^2_g,\bar\sigma_0,\bar\nu,\bar\lambda_1,
\bar\lambda_2$ and $\bar\lambda_3$. For most of the mesons
described by $\Phi$ the exact  relations between the meson
and quark masses become quite involved and need a solution
of the field equation expressing $\la\varphi_q\ra$
in terms of $m_q$.

The case of the flavored pseudoscalars, however, turns out to be
special. For arbitrary values of the parameters $\overline m_g^2,
\bar\sigma_0, \bar\nu, \bar\lambda_1,\bar\lambda_2,\bar\lambda_3$
and arbitrary strength of the sources $j_q$ we find the simple
exact relations
\begin{eqnarray}
\label{9}
\overline M^2_{\pi^\pm}\bar f_\pi&=&\frac{1}{2}
(j_u+j_d)=C(m_u+m_d)\nonumber\\
\overline M^2_{K^\pm}\bar f_{K^\pm}&=&\frac{1}{2}
(j_u+j_s)=C(m_u+m_s)\nonumber\\
\overline M^2_{K^0}\bar f_{K^0}&=&\frac{1}{2}
(j_d+j_s)=C(m_d+m_s).
\end{eqnarray}
These relations are well known in the leading order in chiral
 perturbation theory for $m_q\to 0$ but it may perhaps surprise
that there are no corrections in higher orders in the quark
masses. In fact, a simple exercise in group theory shows
that the  relations (\ref{9}) are exact for an arbitrary form of
the effective potential $U$. Consider first an $SO(N)$ symmetric
theory where  the potential depends on only one vector
$\vec\sigma=(\sigma_1\ldots\sigma_N)$ and
an arbitrary number of singlets $s_k, U=U(\rho,s_k)$. We assume that
$U$ is analytic in $\rho=\frac{1}{2}\vec\sigma^2$ for arbitrary
values of $s_k$ and denote $U'=\partial U/\partial \rho$ etc.
 The $SO(N)$ breaking source is taken in the one-direction such that
the source  term reads $\sigma_1j_1+\sum_k s_k j_k$. The field
equations for $\sigma_a$
\be\label{10}
U'\sigma_a=j_1 \delta_{a1}\ee
admit for $j_1\not= 0$ only the solution
\begin{eqnarray}
\label{11}
\la\sigma_a\ra&=&0\quad{\rm for} \quad a\not= 1\nonumber\\
\la\sigma_1\ra&=&j_1/U'
\end{eqnarray}
 where $U'$ is evaluated at the expectation value for $s_k$ and
$\sigma_a$. Because of the remaining symmetry
$(\sigma_a\to-\sigma_a$ for $a\not= 1$ and $SO(N-1)$ symmetry for
$N\geq 3$) the  mass matrix involves no mixing of the
``Goldstone modes'' $\sigma_{a\not=1}$ with $\sigma_1$
or $s_k$. We can therefore consider the restricted matrix
\be\label{12}
\overline M^2_{ab}=U'\delta_{ab}\quad {\rm for}\quad a,b\not=1.\ee
Comparison with (\ref{11}) yields for the eigenvalues the simple
relation
\be\label{13}
\overline M^2=\frac{j_1}{\la\sigma_1\ra}.\ee

In order to exploit this fact for our case of a
$SU_L(3)\times SU_R(3)$ invariant potential we consider first the
subgroup $SU_L(2)\times SU_R(2)\hat = SO(4)$ acting on the
$u$- and $d$-components. Decomposing $\Phi$ one finds two vectors
$(\pi_1,\pi_2,\pi_3,\sigma_\pi)$ and
$(a_1,a_2,a_3,\eta_a)$ where $\vec\pi$ corresponds to the
 isospin-triplet of the pseudoscalar pions  and $\vec a$ denotes
the isotriplet contained in the scalar $(O^{++})$ octet. The scalar
$\sigma_\pi={\rm Re}(\Phi_{uu}+\Phi_{dd})=\varphi_u+\varphi_d$
and the pseudoscalar $\eta_a={\rm Im}(\Phi_{uu}+\Phi_{dd})$
are isospin singlets. Furthermore, the strange mesons belong to
doublets and the rest are singlets
with respect to $SO(4)$. This $SO(4)$ group is not yet
sufficient for our purpose since both $j_u+j_d$ and
$j_u-j_d$ act as symmetry breaking terms. We will therefore
concentrate on the $SO(3)$ subgroup under which
 $(\sigma_\pi,\pi_1,\pi_2)$ and $(\eta_a, a_1,a_2)$ transform
as vectors. With respect to this subgroup the sources $j_s$
and $j_u-j_d$ are singlets and the only symmetry breaking
term is $j_u+j_d$.  Omitting for a moment the other triplet and
the strange mesons we find precisely the situation described above
and the relation (\ref{13}) becomes equivalent to the first
relation in (\ref{9}). It remains only to be shown that the strange
mesons which belong to two-component spinor representations of
$SO(3)$ and the vector $(\eta_a,a_1,a_2)$ do not disturb this
setting. First we note that for arbitrary $\la\varphi_u\ra,\la\varphi
_d\ra,\la\varphi_s\ra$ the  expectation values of these fields vanish
due to symmetries (strangeness conservation for $K$, electric
charge conservation for $a_1,a_2$, parity for $\eta_a$). They do
therefore not affect the field equations for $(\sigma_\pi,\pi_1,\pi_2)$.
Furthermore, the symmetries forbid any mixing of these fields
with $(\sigma_\pi,\pi_1,\pi_2)$. Therefore the mass matrix for
$(\sigma,\pi_1,\pi_2)$ is not modified by the presence of
these fields either. This establishes the first relation in (\ref{9})
as an exact relation independent of the specific form of $U$ and the
strength of $j_u+j_d$. The two other relations follow immediately
by considering appropriately rotated subgroups which are
obtained from the one discussed above by the substitutions
$(u\leftrightarrow s)$  or $(d\leftrightarrow s)$.

Using (\ref{9}), (\ref{6}) and (\ref{7}) the ratios of current quark
masses can now be inferred from the exact relations
\begin{eqnarray}\label{14}
\frac{m_u+m_s}{m_u+m_d}&=&\frac{M^2_{K^\pm}}{M^2_{\pi^\pm}}
\frac{f_{K^\pm}}{f_{\pi}}\left(\frac{Z_{K^\pm}}{Z_{\pi^\pm}}
\right)^{\frac{1}{2}}\nonumber\\
\frac{m_u+m_s}{m_d+m_s}&=&\frac{M^2_{K^\pm}}{M^2_{K^0}}
\frac{f_{K^\pm}}{f_{K^0}}\left(\frac{Z_{K^\pm}}{Z_{K^0}}
\right)^{\frac{1}{2}}.
\end{eqnarray}
Beyond the electromagnetically corrected meson masses
$M_{\pi^\pm}=135.1\MeV$, $M_{K^0}= 497.7\MeV$,
$M_{K^\pm}=(491.7\pm0.4)\MeV$ (corresponding to $Q=22.7\pm0.8$ in
\cite{3}) these relations involve the decay constants, 
$f_\pi=92.4\MeV$, $f_{K^\pm}=113\MeV$, and ratios of wave function
renormalization constants. Within the linear meson model the isospin
violating ratios $f_{K^\pm} /f_{K^0}$ and $Z_{K^\pm}/Z_{K^0}$ can be
computed \cite{4} as functions of $M^2_i,f_i$ and
$Z_{K^\pm}/Z_{\pi^\pm}$. The ratio $Z_{K^\pm}/Z_{\pi^\pm}$ may
then be related to the mixing in the $\eta-\eta'$-sector and
therefore to the decay constants $f_\eta$ and $f_{\eta'}$. We use from
ref.~\cite{4} the range of values
\begin{eqnarray}\label{15}
  \frac{Z_{K^\pm}}{Z_{\pi^\pm}} &=& 
  0.7085-0.7527\\[2mm]\nonumber
  \frac{Z_{K^\pm}}{Z_{K^0}} &=& 
  (1.00775\pm0.00054)-(1.00657\pm0.00046)\\[2mm]\nonumber
  \frac{f_{K^\pm}}{f_{K^0}} &=&
  (0.99779\pm0.00015)-(0.99725\pm0.00019)\; .
\end{eqnarray}
Here, the errors in parenthesis corresponds to the uncertainty
in the electromagnetically corrected mass
$M_{K^\pm}=(491.7\pm0.4)\MeV$. One finds
\begin{eqnarray}
\label{16}
  \frac{m_u}{m_d} &=& 
  (0.526\pm0.025)-(0.497\pm0.026)\; ,
  \quad [0.533\pm0.043]\nonumber\\[2mm]
  \frac{m_s}{m_d} &=&
  (20.29\pm0.35)-(20.55\pm0.37)\; ,\quad [18.9\pm0.8]\nonumber\\[2mm]
  \frac{m_s}{m_u} &=& 
  (38.60\pm1.17)-(41.4\pm1.4)\; ,\quad [34.4\pm3.7]\; .
\end{eqnarray}
The first two values correspond to the two values of
$Z_{K^\pm}/Z_{\pi^\pm}$  given in (\ref{15}) whereas the error of
each value (given in parenthesis) indicates again the uncertainty
arising from the electromagnetic corrections to the mass difference
$M_{K^0}-M_{K^\pm}$ (same notation as in (\ref{15})).
In square brackets we have also quoted the
results of a recent analysis {}from chiral perturbation theory
\cite{3}. The agreement is satisfactory, with a somewhat lower value
of $m_s$ in chiral perturbation theory. We also note that the
combinations
\begin{eqnarray}
  \label{17}
  \frac{f_{K^\pm}}{f_\pi}\left(\frac{ Z_{K^\pm}}{Z_{\pi^\pm}}
  \right)^{\frac{1}{2}} &=&
  1.03-1.06\nonumber\\[2mm]
  \frac{f_{K^\pm}}{f_{K^0}}\left(\frac{ Z_{K^\pm}}{Z_{K^0}}
  \right)^{\frac{1}{2}} &=&
  (1.00165\pm0.00012)-(1.00053\pm0.00004)
\end{eqnarray}
are very  close to one and corrections to the  leading order
relation  $(m_u+m_s)/(m_u+m_d)=M^2_{K^\pm}/M^2_{\pi^\pm}$
turn therefore out to be small.

For an estimate of the error and for comparison with the results
{}from chiral perturbation theory it is useful to investigate
the ratio
\be\label{18}
\frac{m^2_s-\hat m^2}{m^2_d-m^2_u}=\frac{M^2_K}{M^2_\pi}
\frac{M^2_K-M_\pi}{M^2_{K^0}-M^2_{K^\pm}}(1+\delta_Q)
=Q^2(1+\delta_Q)
\ee 
where $\hat m=(m_u+m_d)/2,\ M^2_K
=(M^2_{K^\pm}+M^2_{K^0})/2,\ f_K= (f_{K^\pm}+f_{K^0})/2,\
Z_K=(Z_{K^\pm}+Z_{K^0})/2, \
M^2_\pi=M^2_{\pi^\pm}$ and (omitting negligible higher order isospin
breaking effects)
\begin{eqnarray}
\label{19}
\delta_Q&=&\frac{f_K}{f_\pi}\left(\frac{Z_K}{Z_\pi}
\right)^{\frac{1}{2}}\left[1+\frac{2(m_s+\hat m)}{m_d-m_u}
\left(1-\frac{f_K}{f_{K^\pm}}\left(\frac{Z_K}{Z_{K^\pm}}
\right)^{\frac{1}{2}}\right)\right]\nonumber\\
&\times&\left[1+\frac{2\hat m}{m_s-\hat m}\left( 1-\frac{f_K}{f_\pi}
\left(\frac{Z_K}{Z_\pi}\right)^{\frac{1}{2}}\right)\right]
^{-1}-1.
\end{eqnarray}
To first order in the quark mass expansion one
has the relations
\be\label{20}
\frac{Z_{K^\pm}-Z_K}{Z_K-Z_\pi}=\frac{\bar f_{K^\pm}
-\bar f_K}{\bar f_K-\bar f_\pi}=-\frac{1}{2}\frac{m_d-m_u} {m_s-\hat
m}\ee and $\delta_Q$ vanishes, consistent with the result from chiral
perturbation theory. Using the values (\ref{15}) quoted from
ref. \cite{4} one finds numerically $\delta_Q\approx 0.11-0.09$. Even
though formally of second order in the quark mass expansion this is a
sizeable correction. It can be explained by the relatively large
deviation of $\bar f_K/\bar f_\pi=1.45-1.41$ from the lowest order
value one. The convergence of the expansion in the strange quark mass
for the coefficients of the isospin violating contributions is
particularly slow \cite{4}.  For fixed $m_u/m_d$ the positive value of
$\delta_Q$ enhances $m_s/\hat m$ as compared to first order chiral
perturbation theory, thus explaining the tendency in
eq. (\ref{16}).

For the second independent
ratio we choose (with $R=(m_s-\hat m)/(m_d-m_u))$
\be\label{21}
\frac{m_s+\hat m}{\hat m}=2\frac{ M^2_K}{M^2_\pi}
\frac{f_K}{f_\pi}\left(\frac{Z_K}{Z_\pi}\right)^{\frac{1}{2}}
=32.9\left(\frac{Z_K}{Z_\pi}\right)^{\frac{1}{2}}=
\frac{2Q^2(1+\delta_Q)}{R}.\ee
The error in this ratio is dominated by the uncertainty
in $Z_K/Z_\pi$. With a rather conservative error of 15\%
for $Z_K/Z_\pi$ we find
\be\label{22}
\frac{m_s}{\hat m}=27.0\pm2.0.\ee
This value turns out slightly higher than the estimate
$24.4\pm1.5$ from chiral perturbation theory \cite{3}.
Our central value corresponds to $R\approx 43$. We observe that
in  contrast to chiral perturbation theory our estimate does not
need any additional assumptions beyond the extraction of the ratio
$Z_K/Z_\pi$ from the two photon decays of $\eta$ and $\eta'$ \cite{4}.
Since this determination is entirely different from the one
used in \cite{3} the agreement of the two estimates is rather
encouraging!

The absolute value of the quark masses needs the constant $C$
in eq. (\ref{9}). Since the current quark masses are normalized
at a given scale (say $\mu=1$ GeV in the $\overline{MS}$ scheme)
the same holds for $C$. Equating the flavor symmetry breaking
term in the quark - and meson - language leads to a relation
for the quark condensate $\la\bar q q\ra$
\be\label{23}
\la \bar q q\ra m_q=-(\la\varphi_q\ra-m_q)j_q.\ee
We use this relation for the up and down quarks and neglect
isospin violation
\be\label{24}
C=-\frac{1}{2}(\la\bar u u\ra+\la\bar d d\ra)
\frac{Z^{\frac{1}{2}}_\pi}{f_\pi-2\hat m Z_\pi^{1/2}}
=(340-410)^2\ {\rm MeV}^2 Z_\pi^{\frac{1}{2}}.
\ee
For the last equation we have taken a standard estimate from sum rules
$\frac{1}{2}(<\bar u u>+<\bar dd>)=-(225\pm25)^3\ {\rm MeV}^3$ and
neglected the correction $\sim\hat mZ^{1/2}_\pi$.  Combining
this with eq. (\ref{9}) yields
\be\label{25}
m_s(1\ {\rm GeV})=(136-198)\ {\rm MeV}\ee The error is dominated by
the uncertainty in the value of the quark condensate. Conversely, any
other independent estimate of $m_u+m_d$ or $m_s$ can be used to fix
$C$ and predict the value of the quark condensate. Recent lattice
estimates \cite{5} seem to favor a value 
$\hat{m}=(2.9\pm0.5){\rm MeV}$.
This would imply
\begin{eqnarray}
  C &=& (545\pm47)^2{\rm MeV}^2 Z_\pi^{1/2}\\[2mm]\nonumber
  \frac{1}{2}<\overline{u}u+\overline{d}d> &=& 
  -(295\pm19)^3{\rm MeV}^3\\[2mm]\nonumber
  m_s &=& (78\pm15){\rm MeV}\; .
\end{eqnarray}

In summary, the quark mass ratios are related in the linear
meson model to the masses and decay constants of the flavored
mesons and their respective wave function renormalization. These
relations are independent of all other parameters of the effective
linear meson model. We use an earlier estimate of the different
wave function renormalizations for $\pi^\pm, K^\pm$ and $K^0$
based on the two photon decay width of the $\eta$ and $\eta'$. This
yields quark mass ratios that resemble very closely the ones
predicted from chiral perturbation theory. The two estimates are
based on entirely independent experimental observations. We also
compute the size of the higher order corrections which are omitted
in present first order estimates from chiral perturbation theory.
They amount typically to an enhancement of around\\
10 \% for $m_s/m_u$ and $m_s/m_d$.


\begin{thebibliography}{12}
\bibitem{1} J. Gasser and H. Leutwyler, Phys. Rep. {\bf 87}
(1982) 77; Nucl. Phys. {\bf B94} (1975) 269; Nucl. Phys.
{\bf B250} (1985) 465
\bibitem{2} S. Weinberg, in {\it A Festschrift for I. I. Rabi},
ed. L. Motz, Trans. New York Acad. Sci. Ser. II 38 (1977) 185
\bibitem{3} H. Leutwyler, preprint hep-ph/9602255
\bibitem{4} D. Jungnickel and C. Wetterich, preprint hep-ph/9606483
\bibitem{5} B.J.~Gough, G.~Hockney, A.X.~El--Khadra, A.S.~Kronfeld,
P.B.~Mackenzie, B.~Mertens, T.~Onogi and J.~Simone, work in
preparation; R.~Gupta and T.~Bhattacharya, LANL preprint
LA--UR--96--1840 (1996) (hep-lat/9605039)
\end{thebibliography}
\end{document}